\documentclass[]{spie}  %>>> use for US letter paper
\usepackage{amsmath,amsfonts,amssymb}
\usepackage{graphicx}
\usepackage[colorlinks=true, allcolors=blue]{hyperref}
\usepackage{wrapfig}

\title{Poisson Flow and Wasserstein Registration of Trees}

\author[a]{Moo K. Chung}
\affil[a]{University of Wisconsin, Madison, WI, USA}
\authorinfo{}

\begin{document} 
\maketitle

\begin{abstract}
Tree-like structures arise in numerous imaging applications, including vascular networks, neuronal arbors, airway trees, and cortical sulcal--gyral folding. We present a nonlinear registration framework based on screened Poisson flow and Wasserstein distance. The screened Poisson equation transforms geometric features into smooth multiscale probability distributions. Registration is formulated by minimizing the Wasserstein distance between these distributions, producing anatomically meaningful correspondences without explicit landmark or branch matching. The framework is demonstrated on the nonlinear registration of cortical sulcal--gyral folding patterns from structural MRI.
\end{abstract}

%\keywords{Tree structures, Heat diffusion, Gaussian mixture, Wasserstein distance, Optimal transport, Nonlinear registration, Sulcal and Gyral patterns}

\section{INTRODUCTION}

Tree-like structures arise in numerous medical imaging applications, including lung vessel trees \cite{chung.2018.EMBCb}, white matter fiber tracts \cite{chung.2011.SPIE} and cortical sulcal--gyral folding patterns \cite{chung.2003.CVPR,chung.2026.EMBC.poisson,chen.2023}. Accurate registration of such structures is fundamental for studying anatomical variability, disease progression, longitudinal change, and population-level morphology. Existing registration methods on trees typically establish correspondences using anatomical landmarks, branch points, centerlines, graph matching, or tree-edit distances
\cite{bullitt.2003,feragen.2015}. Although successful in specific settings, these approaches often require explicit branch correspondence and can become unstable under substantial geometric variability, missing branches, or differences in topology.

Optimal transport has emerged as a powerful mathematical framework for comparing geometric objects by measuring the minimum cost required to transport one probability distribution into another
\cite{peyre.2019,song.2023}. Unlike pointwise matching methods, Wasserstein distance compares entire distributions while preserving their global spatial organization, making it naturally suited for nonlinear registration problems. However, applying optimal transport to tree structures requires a smooth probabilistic representation that faithfully captures the underlying branching geometry while remaining robust to local geometric irregularities and discretization noise. Thus, we introduce a screened Poisson representation of tree structures. Solving a screened Poisson equation transforms a discrete tree measure into a smooth multiscale probability distribution. Through the resolvent identity, the solution is an exponentially weighted superposition of heat-kernel diffusions and therefore admits a multiscale Gaussian-mixture interpretation\cite{chung.2026.EMBC.poisson}. Registration is then formulated as the minimization of the Wasserstein distance between the resulting screened Poisson representations, producing smooth nonlinear correspondences without requiring explicit landmark or branch matching.

\begin{wrapfigure}{r}{0.4\textwidth}
\vspace{0.25cm}
\centering
\includegraphics[width=\linewidth]{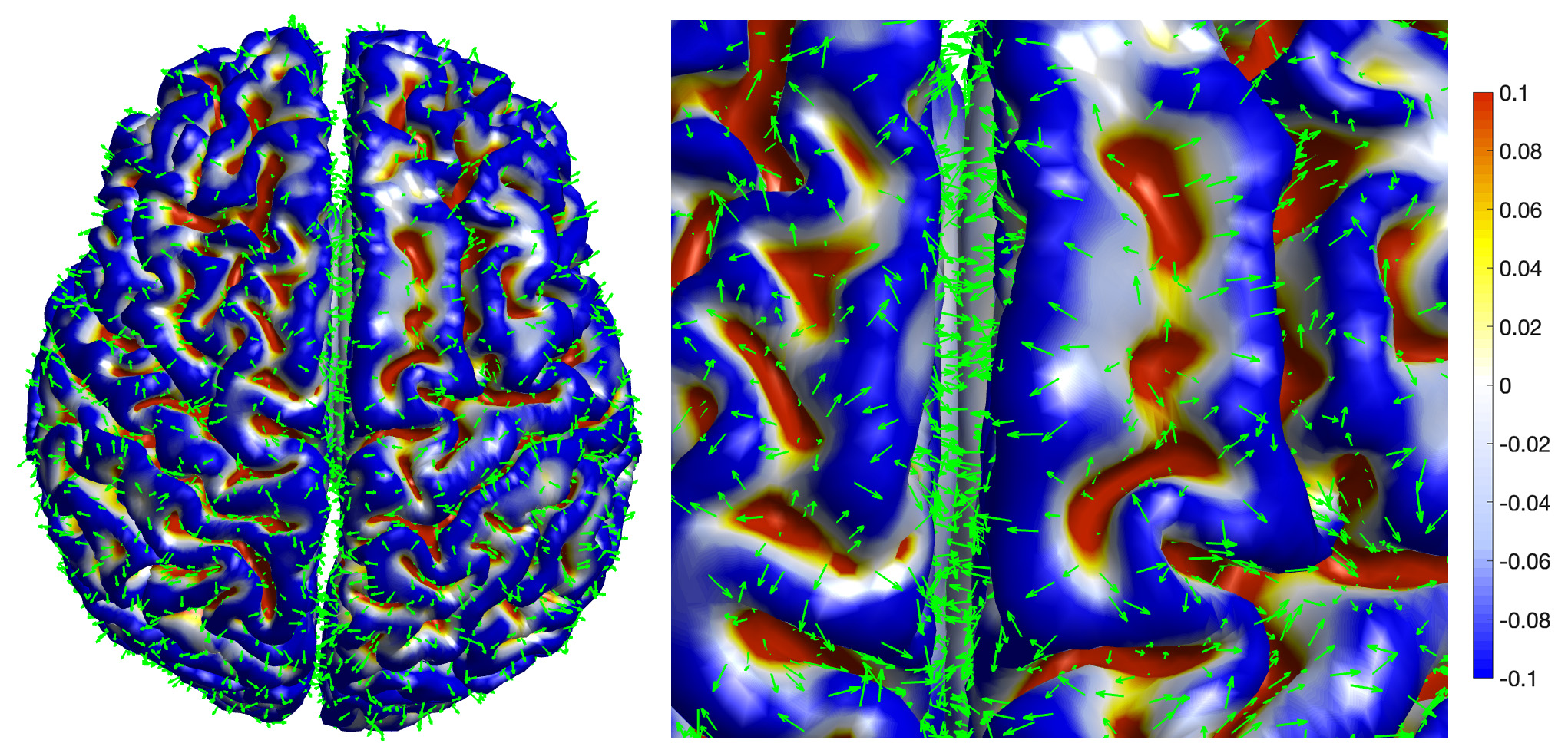}
\caption{Sulcal patterns (red), extracted as the positive part of the screened Poisson solution computed from the normalized mean-curvature field, form tree-like structures on the cortical surface. Green arrows denote the Poisson flow, defined as the negative gradient of the screened Poisson potential. During registration, this flow defines the velocity field that iteratively drives the spherical deformation toward the target sulcal pattern.}
\label{fig:sulcal}
\vspace{0.7cm}
\end{wrapfigure}

The main contributions of this work are threefold. (i) We introduce a screened Poisson representation that transforms tree structures into smooth multiscale probability distributions. (ii) We formulate nonlinear tree registration as an optimal transport problem by minimizing the Wasserstein distance between these probabilistic representations, eliminating the need for explicit landmark or branch correspondence. (iii) We demonstrate the proposed framework on cortical sulcal patterns extracted from structural MRI (Figure \ref{fig:sulcal}) \cite{huang.2019.MICCAI,huang.2020.TMI}. Cortical folding presents a particularly challenging application because of substantial inter-subject variability in folding geometry and the absence of anatomically meaningful pointwise correspondences beyond conventional surface registration \cite{fischl.2012,im.2019,lyu.2018.MIA}. The proposed method achieves smooth nonlinear alignment of complex sulcal folding patterns without requiring explicit anatomical landmarks or branch correspondences.

\section{Screened Poisson Representation of Trees}
Consider a tree structure embedded in a Riemannian manifold $\mathcal M$ and represented by a probability measure
$
\mu
=
\sum_{i=1}^{m}
a_i\delta_{p_i},
$
where $\delta_{p_i}$ denotes the Dirac-delta function at node $p_i$, $a_i\ge0$, and
$\sum_i a_i=1$ (Figure \ref{fig:treetoy}). The underlying graph defined by the nodes determines the tree topology. The screened Poisson representation of the tree is then obtained by solving
\hspace*{\fill}
\[
(\mathcal L+\lambda I)u=\mu,
\]
where $\mathcal L$ is the Laplace--Beltrami operator and $\lambda>0$ is a regularization parameter \cite{chung.2003.CVPR,chung.2026.EMBC.poisson,chen.2023}. 
The inverse operator satisfies the resolvent identity
\[
u
=
(\mathcal L+\lambda I)^{-1}\mu
=
\int_0^\infty
e^{-\lambda t}
H_t\mu\,dt,
\]
where $H_t=e^{-t\mathcal L}$ is the heat diffusion operator at diffusion time $t$. Thus, the screened Poisson solution is an exponentially weighted superposition of heat-kernel diffusions. Thus, the screened Poisson solution is an exponentially weighted superposition of heat-kernel diffusions, yielding a multiscale Gaussian mixture representation of the underlying tree structure\cite{chung.2026.EMBC.poisson}. Since
$
\int_{\mathcal M}u\,dV
=
\frac{1}{\lambda},
$
we normalize the screened Poisson representation as
\[
p_\lambda^\mu
=
\lambda(\mathcal L+\lambda I)^{-1}\mu
=
\int_0^\infty
\lambda e^{-\lambda t}
H_t\mu\,dt,
\]
which makes $p_\lambda^\mu$ as a probability measure whenever $\mu$ is.

\section{Wasserstein Distance between Trees}
Given two probability measures $\mu$ and $\nu$ on $\mathcal M$, the 2-Wasserstein 
 \hspace*{\fill}
\begin{wrapfigure}{r}{0.35\textwidth}
\vspace{-0.75cm}
\centering
\includegraphics[width=\linewidth]{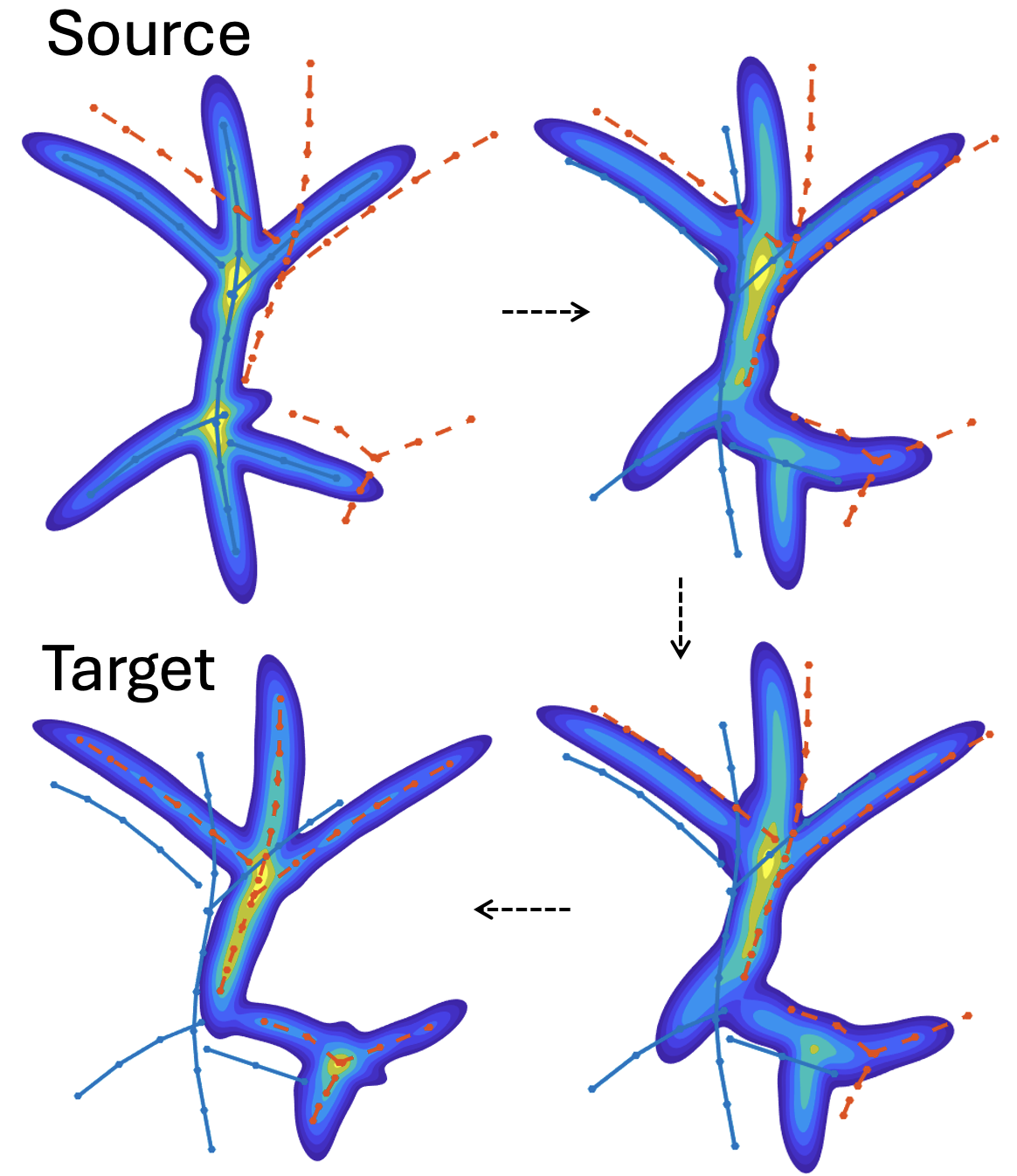}
\caption{Wasserstein geodesic between screened Poisson representations of two tree structures with different topologies. The source tree (blue) is connected, whereas the target tree (red) contains disconnected components and different branching patterns. The intermediate panels illustrate smooth nonlinear deformation between the two structures.}
\label{fig:treetoy}
\vspace{-1cm}
\end{wrapfigure}
distance is defined as
$$
W_2^2(\mu,\nu)
=
\inf_{\gamma\in\Pi(\mu,\nu)}
\int_{\mathcal M\times\mathcal M}
d(x,y)^2\,d\gamma(x,y),
$$
where $d(x,y)$ is the geodesic distance on $\mathcal M$, $\Pi(\mu,\nu)$ denotes the set of all joint distributions having marginals $\mu$ and $\nu$, and $\gamma$ specifies the amount of probability mass transported from $x$ to $y$ (Figure \ref{fig:treetoy}). Then the Wasserstein distance between the screened Poisson representations of two tree structure is 
$
W_2\!\left(
p_\lambda^\mu,
p_\lambda^\nu
\right),
$
which admits no closed-form expression because each Poisson representation is a continuous mixture of Gaussian kernels over both source locations and diffusion times. Numerical computation is therefore required. However, the Poisson representation inherits several important theoretical properties from heat diffusion. 

\noindent\textbf{Theorem 1.}
Let $\mathcal{M}$ be a complete Riemannian manifold with nonnegative Ricci curvature, and let $\mu$ and $\nu$ be probability measures with finite second moments. Then $
W_2\!\left(
p_\lambda^\mu,p_\lambda^\nu
\right)
\le
W_2(\mu,\nu).
$

\noindent\textit{Proof.}
For each $t\geq0$, let $\gamma_t$ be an optimal coupling between $H_t\mu$ and $H_t\nu$. Their exponentially weighted mixture
$
\overline{\gamma}_\lambda
=
\int_0^\infty
\lambda e^{-\lambda t}\gamma_t\,dt
$
is a coupling between $p_\lambda^\mu$ and $p_\lambda^\nu$. Therefore,
\[
\begin{aligned}
W_2^2(p_\lambda^\mu,p_\lambda^\nu)
&\le
\int_0^\infty
\lambda e^{-\lambda t}
W_2^2(H_t\mu,H_t\nu)\,dt  \\
&\le
\int_0^\infty
\lambda e^{-\lambda t}
W_2^2(\mu,\nu)\,dt =
W_2^2(\mu,\nu),
\end{aligned}
\]
where the second inequality follows from the Wasserstein contraction of heat flow under nonnegative Ricci curvature \cite{von.2005}. Taking square roots completes the proof.
\hfill$\square$

Theorem~1 shows that the screened Poisson representation is contractive in the Wasserstein metric. Consequently, replacing a tree by its screened Poisson representation cannot increase the Wasserstein distance between trees. This provides a stable multiscale representation that suppresses small-scale geometric variability while preserving global structural differences, making it well suited for nonlinear registration.

\begin{figure}
\centering
\includegraphics[width=0.9\linewidth]{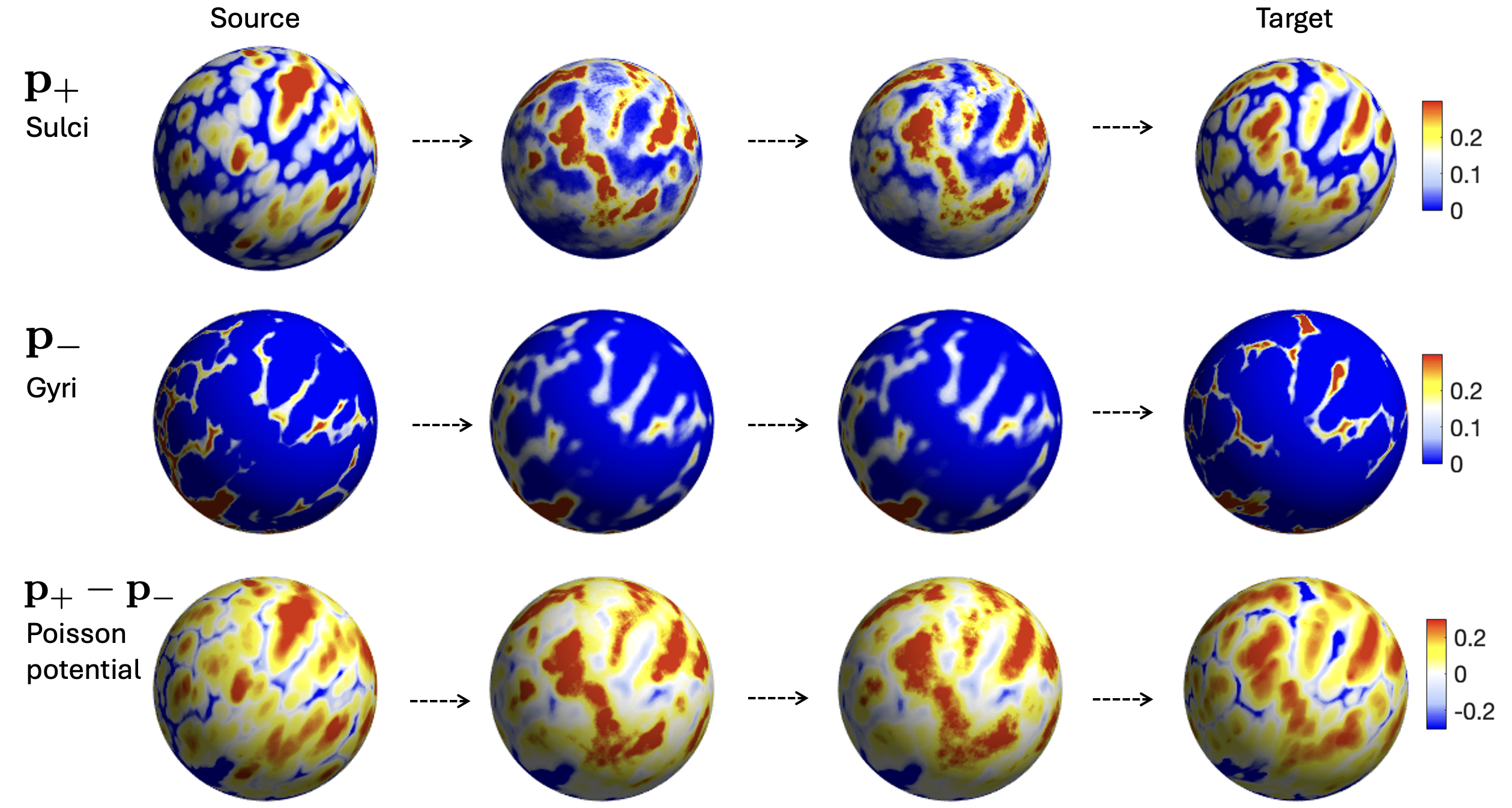}
\caption{Optimal transport of cortical folding patterns. The top row shows the sulcal probability ${\bf p}_{+}$, the middle row the gyral probability ${\bf p}_{-}$, and the bottom row the signed density ${\bf p}_{+}-{\bf p}_{-}$. From left to right are the source, two intermediate iterations, and the target. At each iteration, a screened Poisson equation is solved from the residual, and its gradient updates the deformation toward the target.}
\label{fig:tree}
\end{figure}

\section{Application to Cortical Sulcal Pattern}

For each subject, mean curvature is computed on the cortical surface mesh using local quadratic surface fitting \cite{chung.2003.CVPR}. The Laplace--Beltrami operator is discretized using the cotangent finite-element formulation \cite{chung.2003.CVPR,huang.2020.TMI}. Let ${\bf L}$ and ${\bf A}$ denote the stiffness and mass matrices. Given the discrete mean-curvature field ${\bf h}$, the screened Poisson potential is computed by solving
 \hspace*{\fill}
\begin{wrapfigure}{r}{0.3\textwidth}
\vspace{-.75cm}
\centering
\includegraphics[width=\linewidth]{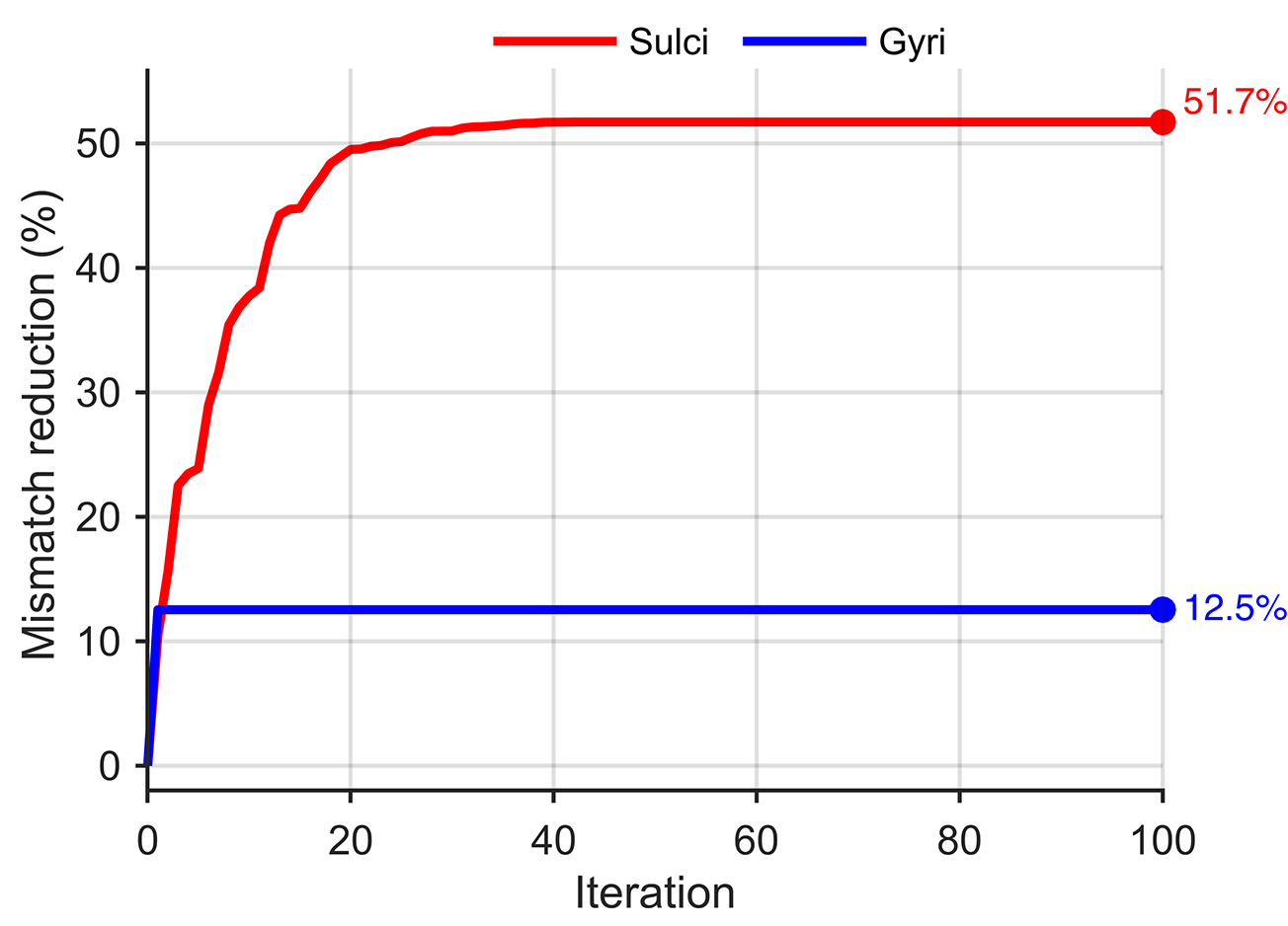}
\caption{Reduction of the normalized sulcal and gyral probability mismatch during screened Poisson-flow registration. The mismatch quickly decreases monotonically over successive iterations, demonstrating convergence of the proposed registration algorithm.}
\label{fig:plot}
\vspace{0.7cm}
\end{wrapfigure}
\[
({\bf L}+\lambda{\bf A}){\bf u}
=
{\bf A}{\bf h}.
\]
The signed potential is decomposed into positive and negative components,
\[
{\bf u}_{+}
=
\max\{{\bf u},0\},
\qquad
{\bf u}_{-}
=
\max\{-{\bf u},0\},
\]
representing sulcal and gyral folding, respectively. Each component is normalized with respect to the finite-element area measure,
\[
{\bf p}_{+}
=
\frac{{\bf u}_{+}}
{{\bf 1}^{\top}{\bf A}{\bf u}_{+}},
\qquad
{\bf p}_{-}
=
\frac{{\bf u}_{-}}
{{\bf 1}^{\top}{\bf A}{\bf u}_{-}},
\]
yielding sulcal and gyral probability densities.

The sulcal and gyral probability densities are registered independently using the same iterative registration algorithm (Figure~\ref{fig:tree}). Starting from the identity map, the source probability mass is transported toward the target density. At each iteration, the transported density is compared with the target density to compute a residual field, which serves as the source term of a screened Poisson equation on the sphere. The tangential gradient of the resulting potential defines the transport direction, which is normalized to obtain a smooth velocity field. The spherical map is updated through the Riemannian exponential map, while a backtracking line search adaptively reduces the step size until the density mismatch decreases monotonically \cite{nocedal.2006}. The iteration continues until convergence of the density mismatch (Figure~\ref{fig:plot}). We achieved up to a 51.7\% reduction in sulcal mismatch, measured by the $\ell_2$-distance between the transported and target sulcal probability densities.

\section{Cognitive and Genetic Applications}

The final manuscript will demonstrate the proposed registration framework on a structural MRI cohort of 400 subjects comprising monozygotic and dizygotic twin pairs. Following registration, pairwise similarities between cortical sulcal patterns will be analyzed to investigate associations with cognitive performance and genetic relatedness. These applications demonstrate the potential of the proposed Poisson-flow representation for population studies of cortical folding.

% References
\bibliography{reference.2026.07.31} % bibliography data in report.bib

\begin{thebibliography}{10}

\bibitem{chung.2018.EMBCb}
Chung, M., Wang, Y., and Wu, G., ``Heat kernel smoothing in irregular image
  domains,'' {\em International Conference of the IEEE Engineering in Medicine
  and Biology Society (EMBC)} ,  5101--5104 (2018).

\bibitem{chung.2011.SPIE}
Chung, M., Adluru, N., Dalton, K., Alexander, A., and Davidson, R., ``Scalable
  brain network construction on white matter fibers,'' in [{\em Proc. of
  SPIE}{\nolinebreak\hspace{0.1em}]},   {\bf 7962},  79624G (2011).

\bibitem{chung.2003.CVPR}
Chung, M., Worsley, K., Robbins, S., and Evans, A., ``Tensor-based brain
  surface modeling and analysis,'' in [{\em IEEE Conference on Computer Vision
  and Pattern Recognition (CVPR)}{\nolinebreak\hspace{0.1em}]},   {\bf I},
  467--473 (2003).

\bibitem{chung.2026.EMBC.poisson}
Chung, M., Maccotta, L., and Struck, A., ``Poisson flow model of cortical
  folding pattern,'' in [{\em Proceedings of the Annual International
  Conference of the IEEE Engineering in Medicine and Biology Society
  (EMBC)}{\nolinebreak\hspace{0.1em}]},   1--5 (2026).

\bibitem{chen.2023}
Chen, Z., Das, S., and Chung, M., ``Sulcal pattern matching with the
  wasserstein distance,'' in [{\em 2023 IEEE 20th International Symposium on
  Biomedical Imaging (ISBI)}{\nolinebreak\hspace{0.1em}]},   1--5, IEEE (2023).

\bibitem{bullitt.2003}
Bullitt, E., Gerig, G., Pizer, S., Lin, W., and Aylward, S., ``Measuring
  tortuosity of the intracerebral vasculature from {MRA} images,'' {\em IEEE
  transactions on medical imaging}~{\bf 22},  1163--1171 (2003).

\bibitem{feragen.2015}
Feragen, A., Lauze, F., and Hauberg, S., ``Geodesic exponential kernels: When
  curvature and linearity conflict,'' in [{\em Proceedings of the IEEE
  conference on computer vision and pattern
  recognition}{\nolinebreak\hspace{0.1em}]},   3032--3042 (2015).

\bibitem{peyre.2019}
Peyr{\'e}, G. and Cuturi, M., ``Computational optimal transport: {W}ith
  applications to data science,'' {\em Foundations and Trends in Machine
  Learning}~{\bf 11}(5-6),  355--607 (2019).

\bibitem{song.2023}
Songdechakraiwut, T. and Chung, M., ``Topological learning for brain
  networks,'' {\em Annals of Applied Statistics}~{\bf 17},  403--433 (2023).

\bibitem{song.2021.MICCAI}
Songdechakraiwut, T., Shen, L., and Chung, M., ``Topological learning and its
  application to multimodal brain network integration,'' {\em Medical Image
  Computing and Computer Assisted Intervention (MICCAI)}~{\bf 12902},  166--176
  (2021).

\bibitem{huang.2019.MICCAI}
Huang, S.-G., Lyu, I., Qiu, A., and Chung, M., ``Fast polynomial approximation
  to heat diffusion in manifolds,'' {\em MICCAI}~{\bf 11767},  48--56 (2019).

\bibitem{fischl.2012}
Fischl, B., ``{FreeSurfer},'' {\em NeuroImage}~{\bf 62},  774--781 (2012).

\bibitem{im.2019}
Im, K. and Grant, P., ``Sulcal pits and patterns in developing human brains,''
  {\em Neuroimage}~{\bf 185},  881--890 (2019).

\bibitem{lyu.2018.MIA}
Lyu, I., Kim, S., Girault, J., Gilmore, J., and Styner, M., ``A cortical
  shape-adaptive approach to local gyrification index,'' {\em Medical image
  analysis}~{\bf 48},  244--258 (2018).

\bibitem{huang.2020.TMI}
Huang, S.-G., Lyu, I., Qiu, A., and Chung, M., ``Fast polynomial approximation
  of heat kernel convolution on manifolds and its application to brain sulcal
  and gyral graph pattern analysis,'' {\em IEEE Transactions on Medical
  Imaging}~{\bf 39},  2201--2212 (2020).

\bibitem{von.2005}
von Renesse, M.-K. and Sturm, K.-T., ``Transport inequalities, gradient
  estimates, entropy and {R}icci curvature,'' {\em Communications on Pure and
  Applied Mathematics}~{\bf 58},  923--940 (2005).

\bibitem{nocedal.2006}
Nocedal, J. and Wright, S.,  [{\em Numerical
  Optimization}{\nolinebreak\hspace{0.1em}]}, Springer (2006).

\end{thebibliography}
\bibliographystyle{spiebib} % makes bibtex use spiebib.bst

\end{document}